\documentclass{article}
\pdfoutput=1

\usepackage[english]{babel}
\usepackage[utf8]{inputenc}
\usepackage{textcomp}
\usepackage[letterpaper,top=2cm,bottom=2cm,left=3cm,right=3cm,marginparwidth=1.75cm]{geometry}
\usepackage{array}
\usepackage{caption}
\usepackage{amsmath}
\usepackage{graphicx}
\usepackage[colorlinks=true, allcolors=blue]{hyperref}
\usepackage{subcaption}

\title{MedImageInsight for Thoracic Cavity Health Classification from Chest X-rays}
\author{
Rama Krishna Boya, Mohan Kireeti Magalanadu, Azaruddin Palavalli, Rupa Ganesh Tekuri,\\
Amrit Pattanayak, Prasanthi Enuga, Vignesh Esakki Muthu, Vivek Aditya Boya \\
DeepInfinity Ltd, London, United Kingdom
}

\begin{document}
\maketitle

\begin{abstract}
Chest radiography remains one of the most widely used imaging modalities for thoracic diagnosis, yet increasing imaging volumes and radiologist workload continue to challenge timely interpretation. In this work, we investigate the use of MedImageInsight, a medical imaging foundational model, for automated binary classification of chest X-rays into Normal and Abnormal categories. Two approaches were evaluated: (1) fine-tuning MedImageInsight for end-to-end classification, and (2) employing the model as a feature extractor for a transfer learning pipeline using traditional machine learning classifiers. Experiments were conducted using a combination of the ChestX-ray14 dataset and real-world clinical data sourced from partner hospitals. The fine-tuned classifier achieved the highest performance, with an ROC-AUC of 0.888 and superior calibration compared to the transfer learning models, demonstrating performance comparable to established architectures such as CheXNet. These results highlight the effectiveness of foundational medical imaging models in reducing task-specific training requirements while maintaining diagnostic reliability. The system is designed for integration into web-based and hospital PACS workflows to support triage and reduce radiologist burden. Future work will extend the model to multi-label pathology classification to provide preliminary diagnostic interpretation in clinical environments.
\end{abstract}

\section{Introduction}
Radiologists face an increasing burden, with rising imaging volumes over the past 15 years and burnout affecting diagnostic accuracy [1]. These pressures have highlighted the need for support systems in medical imaging [2]. Artificial Intelligence (AI) can be a powerful tool for assisting in the diagnosis of thoracic cavities. CheXNet [3] and other DeepChest Models [4] focusing on multi-disease detection show promising results, with performance comparable to that of expert radiologists. AI aids interpretation and prioritisation of critical cases and has the potential to reduce diagnostic errors, speed up workflows, and ease the burden on radiologists [5]. \newline \newline
Traditional thoracic cavity diagnosis methods relied on classical machine learning algorithms using handcrafted image features, but these approaches often lacked sensitivity and struggled with generalisation across varied clinical settings [7]. The introduction of deep learning, particularly Convolutional Neural Networks (CNNs), brought significant improvements by automatically learning features and achieving high performance in disease classification from chest X-rays. However, CNN-based systems are often limited in flexibility, typically trained for narrow tasks with labelled data. To overcome these limitations, recent advancements in foundational models have led to the development of more general-purpose AI systems capable of multimodal learning [8-11]. These models, such as MedImageInsight, Microsoft’s foundation model for health, enable tasks such as zero-shot classification and image-to-text generation, offering scalable, adaptable solutions for real-world thoracic diagnosis [6]. \newline \newline
Unlike previous work that primarily evaluates foundational models on public datasets, our study conducts a comprehensive comparison of fine-tuning and transfer-learning approaches using MedImageInsight on both ChestX-ray14 and real-world, multi-institution clinical data. This combined evaluation demonstrates how a healthcare foundation model can be effectively adapted for scalable, PACS-integrated triage in practical hospital environments. This study explores the development and performance of two classifiers constructed using MedImageInsight for medical image interpretation. Binary classifiers for categorising chest X-rays as Normal or Abnormal were developed: one by fine-tuning the model and the other using a transfer learning approach [12]. In summary, the benefits of this work are as follows:
\begin{enumerate}
    \item The planned integration of the model into web-based applications and hospital systems aims to provide healthcare professionals with instant, interpretable diagnostic outputs—detecting abnormalities in scans to prioritise high-risk patients and alerting physicians to next steps, such as further screening or specialist consultations. 
    \item By embedding this AI system into existing workflows, the solution becomes accessible, scalable, and directly applicable in real-world healthcare settings. This contribution not only enhances diagnostic accuracy but also supports digital health transformation by providing AI-driven insights and enabling effective time management.
\end{enumerate}
We believe that the findings of this study have the potential to substantially reduce the workload of healthcare professionals and assist in addressing the ethical dilemma of scan prioritisation through the use of these models. In future work, we aim to develop a model capable of providing preliminary diagnostic findings for any chest X-ray.

\section{Results}
\subsection{Transfer Learning approach}
The performance of various Deep Learning (DL) classifiers trained on embeddings generated by MedImageInsight and classifier fine-tuned from MedImageInsight is compared in Table 1. The classifiers evaluated include K-Nearest Neighbours (KNN), Logistic Regression (LogReg), Support Vector Machine (SVM), Random Forest (RF), and Multilayer Perceptron (MLP).
\begin{table}[htbp]
\centering
\scriptsize %
\caption{Comparison of performance metrics across different classification models}
\label{tab:model_comparison}
\begin{tabular}{lcccccc}
\hline
\textbf{Metric} & \textbf{KNN} & \textbf{LogReg} & \textbf{SVM} & \textbf{RF} & \textbf{MLP} & \textbf{Fine-tuned Classifier} \\
\hline
Accuracy      & 0.7125 & 0.7499 & 0.7621 & 0.7558 & 0.7490 & 0.7860 \\
Precision     & 0.6532 & 0.6944 & 0.7239 & 0.7204 & 0.7469 & 0.8750 \\
Recall        & 0.7340 & 0.7673 & 0.7400 & 0.7240 & 0.6468 & 0.7230 \\
F1-Score      & 0.6913 & 0.7291 & 0.7318 & 0.7222 & 0.6933 & 0.7920 \\
ROC AUC       & 0.7602 & 0.8200 & 0.8277 & 0.8182 & 0.8038 & 0.8880 \\
Brier Score   & 0.2200 & 0.2070 & 0.1680 & 0.1760 & 0.1810 & 0.1370 \\
\hline
\end{tabular}
\end{table}

\begin{figure}[htbp]
\centering
\includegraphics[width=\textwidth]{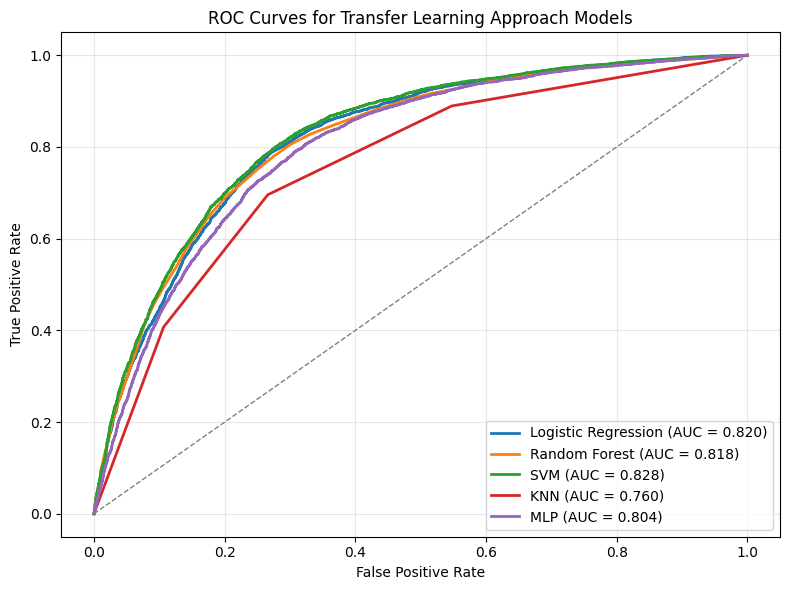}
\caption{Receiver Operating Characteristic (ROC) curves for the evaluated classifiers (SVM, Random Forest, MLP, Logistic Regression, and KNN) in the transfer learning approach.}
\label{fig:transfer_learning_roc}
\end{figure}

A sample of 10,000 images from the ChestX-ray14 dataset was used to train and evaluate these models. ROC curves of each classifier for each class from this approach are shown in Fig. 1. With the results presented, SVM can be the classifier that achieves the best performance.

\subsection{Fine Tuning}
The performance of the classifier built for fine-tuning MedImageInsight is shown in Table 1, compared with other classifiers in the transfer learning approach. An ROC curve is shown in Fig. 4a. In Figs. 2 and 3, Precision-Recall curves and Calibration plots of both approaches are shown.

\begin{figure}[htbp]
    \centering
    \begin{subfigure}[b]{0.48\textwidth}
        \centering
        \includegraphics[width=\linewidth, height=8cm, keepaspectratio]{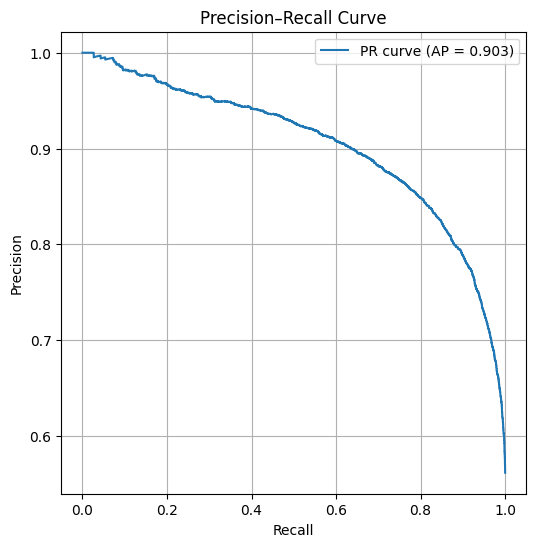}
        \label{fig:precision_curve_finetuned}
    \end{subfigure}
    \hfill
    \begin{subfigure}[b]{0.48\textwidth}
        \centering
        \includegraphics[width=\linewidth, height=10cm, keepaspectratio]{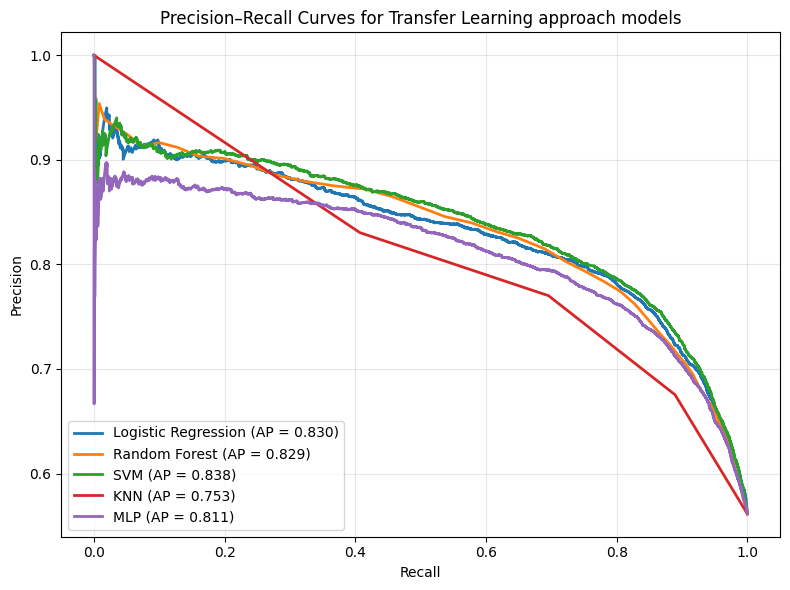}
        \label{fig:precision_curve_transfer}
    \end{subfigure}
    \caption{Precision-Recall curves of the finetuned classifier (a) (on the left) and other classifiers of the transfer learning approach (b) (on the right).}
    \label{fig:precision_recall_comparison}
\end{figure}

\begin{figure}[htbp]
    \centering
    \begin{subfigure}[b]{0.48\textwidth}
        \centering
        \includegraphics[width=\linewidth, height=8cm, keepaspectratio]{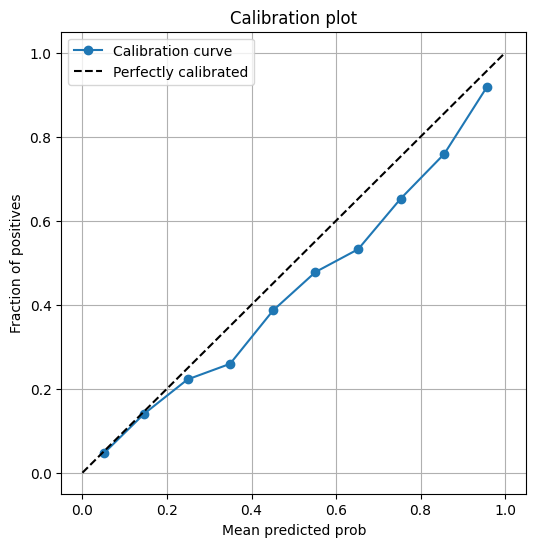}
        \label{fig:calibration_finetuned}
    \end{subfigure}
    \hfill
    \begin{subfigure}[b]{0.48\textwidth}
        \centering
        \includegraphics[width=\linewidth, height=10cm, keepaspectratio]{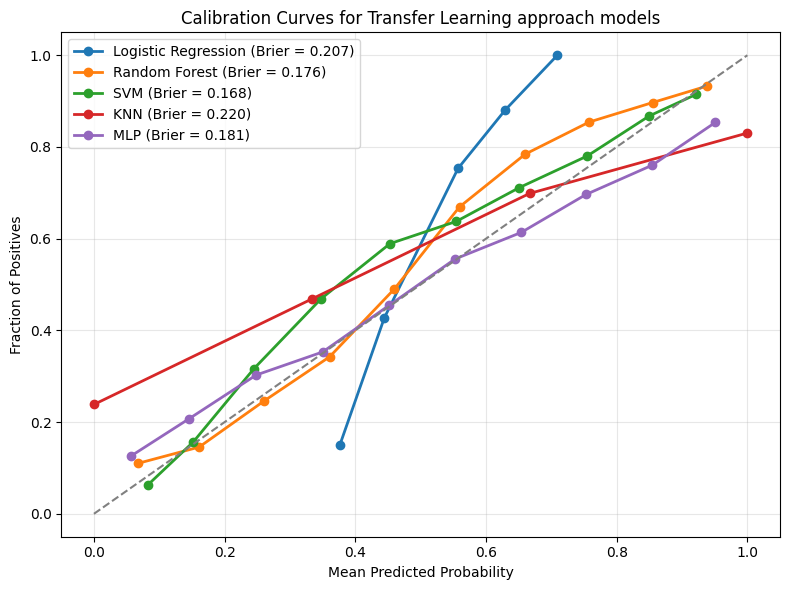}
        \label{fig:calibration_transfer}
    \end{subfigure}
   
    \caption{Calibration plots of the finetuned classifier (a)(on the left) and other classifiers of the transfer learning approach (b) (on the right).}
    \label{fig:calibration_comparison}
\end{figure}

The results show that the classifier from the fine-tuning approach is performing significantly better. With an ROC of 0.888, this model is comparable to other SOTA chest X-Ray classifiers. The ROC curves are plotted against other ROC curves and against known Chest X-Ray classifiers, such as the architecture by Wong et al. [14] and Rajpurkar et al. (CheXNet) [3], in Fig. 4.

\begin{figure*}[htbp]
    \centering
    \begin{subfigure}[b]{0.48\textwidth}
        \centering
        \includegraphics[width=\linewidth, height=5cm]{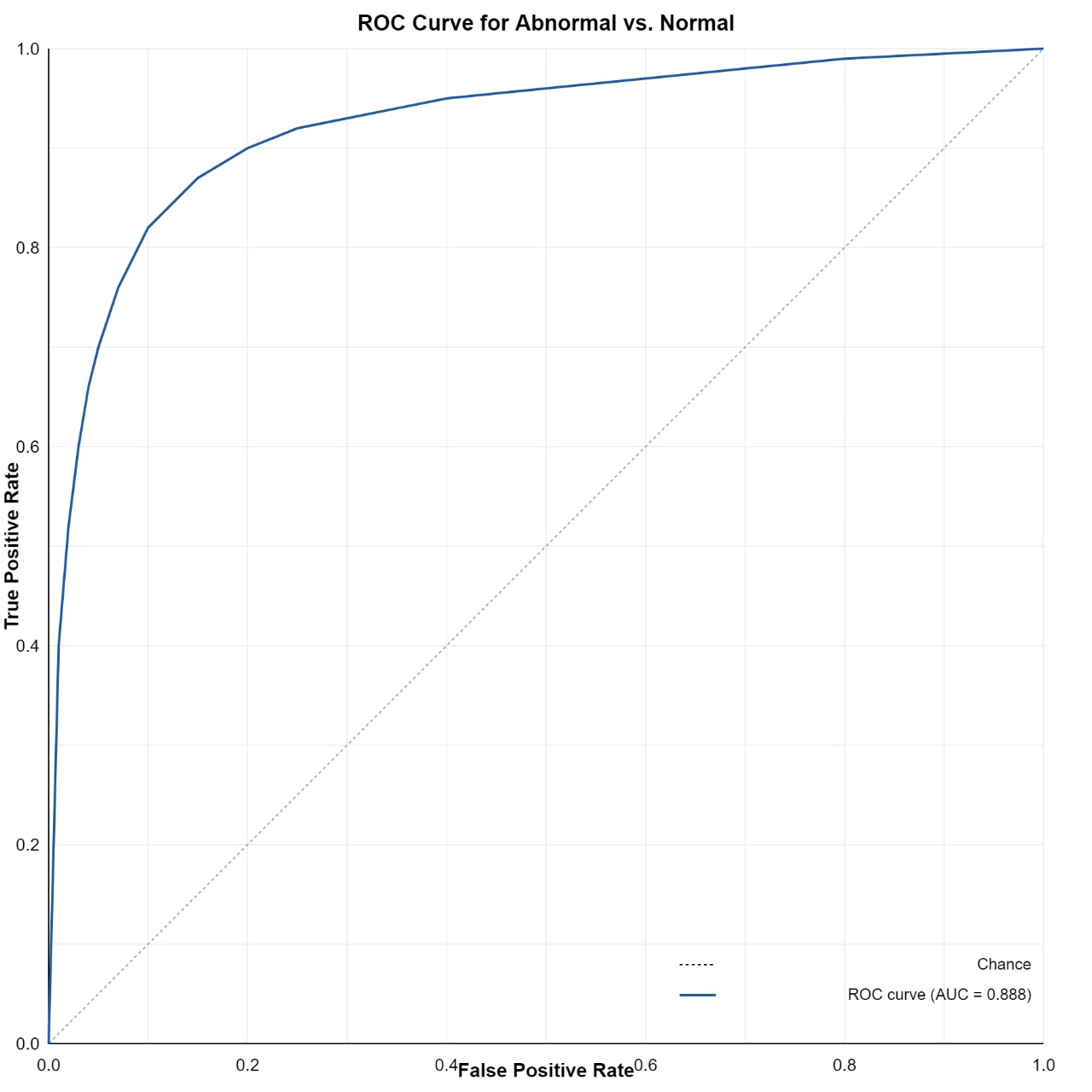}
        \caption{ROC curve for our model.}
        \label{fig:roc_diai}
    \end{subfigure}
    \hfill
    \begin{subfigure}[b]{0.48\textwidth}
        \centering
        \includegraphics[width=\linewidth, height=5cm]{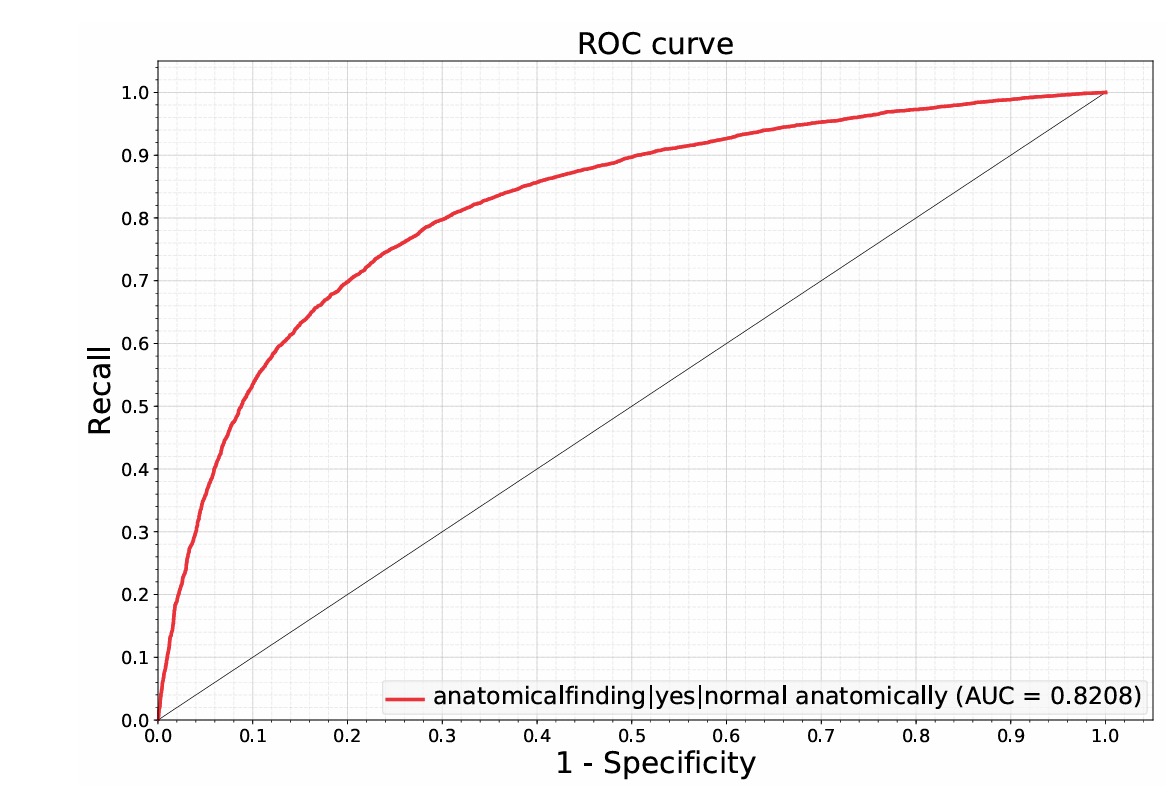}
        \caption{ROC curve for architecture by Wong et al. [14].}
        \label{fig:roc_normal}
    \end{subfigure}
    \hfill
    \begin{subfigure}[b]{0.48\textwidth}
        \centering
        \includegraphics[width=\linewidth, height=5cm]{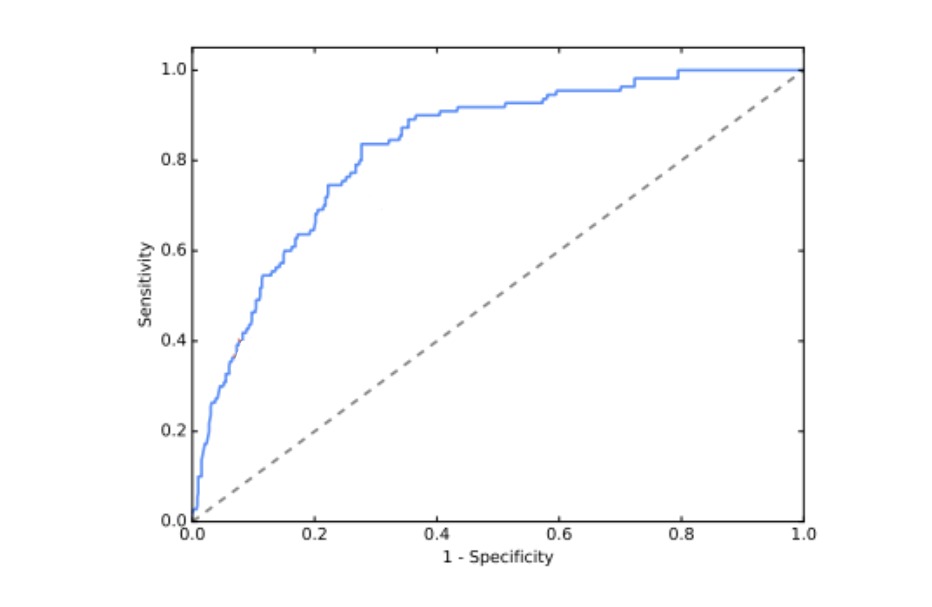}
        \caption{ROC curve for CheXNet model by Rajpurkar et al. [3]}
        \label{fig:roc_chexnet}
    \end{subfigure}
    \caption{Comparison of ROC curves for three models: (a) (on top left) Our model, (b) (on top right) Wong et al. [14], and (c) (at bottom) CheXNet by Rajpurkar et al. [3]. All curves show the trade-off between sensitivity (true positive rate) and 1 specificity (false positive rate).}
    \label{fig:roc_comparison}
\end{figure*}
\section{Methods}
\subsection{Training}
MedImageInsight is a Healthcare Foundational Model developed by Microsoft for medical imaging analysis. It employs self-supervised and multimodal learning techniques to generate high-quality image embeddings from radiological images. The model has been trained on extensive, domain-specific datasets and is optimised for downstream applications such as image classification, anomaly detection, and clinical report generation. By providing generalised, transferable image representations, MedImageInsight enables efficient model development with limited task-specific data through fine-tuning or transfer learning. \newline \newline
In the first approach, the pretrained MedImageInsight model was fine-tuned for binary classification of chest X-ray images. The objective of this model was to distinguish between Normal and Abnormal cases. By adapting the foundational model through fine-tuning on a task-specific dataset, the classifier was optimised to learn discriminative radiological features indicative of abnormal findings, thereby enhancing its clinical relevance and diagnostic accuracy. \newline \newline
In the second approach, a transfer learning strategy was employed using embeddings extracted from MedImageInsight. These embeddings were used to train a downstream classifier to distinguish between Normal and Abnormal cases, as in the first one. Five Deep Learning (DL) classifiers were used to identify the best-performing classifier and compare it with the previous approach.  The classifiers include K-Nearest Neighbours (KNN), Logistic Regression (LogReg), Support Vector Machine (SVM), Random Forest (RF), and Multilayer Perceptron (MLP). 
\begin{figure}[htbp]
    \centering
    \includegraphics[width=\textwidth]{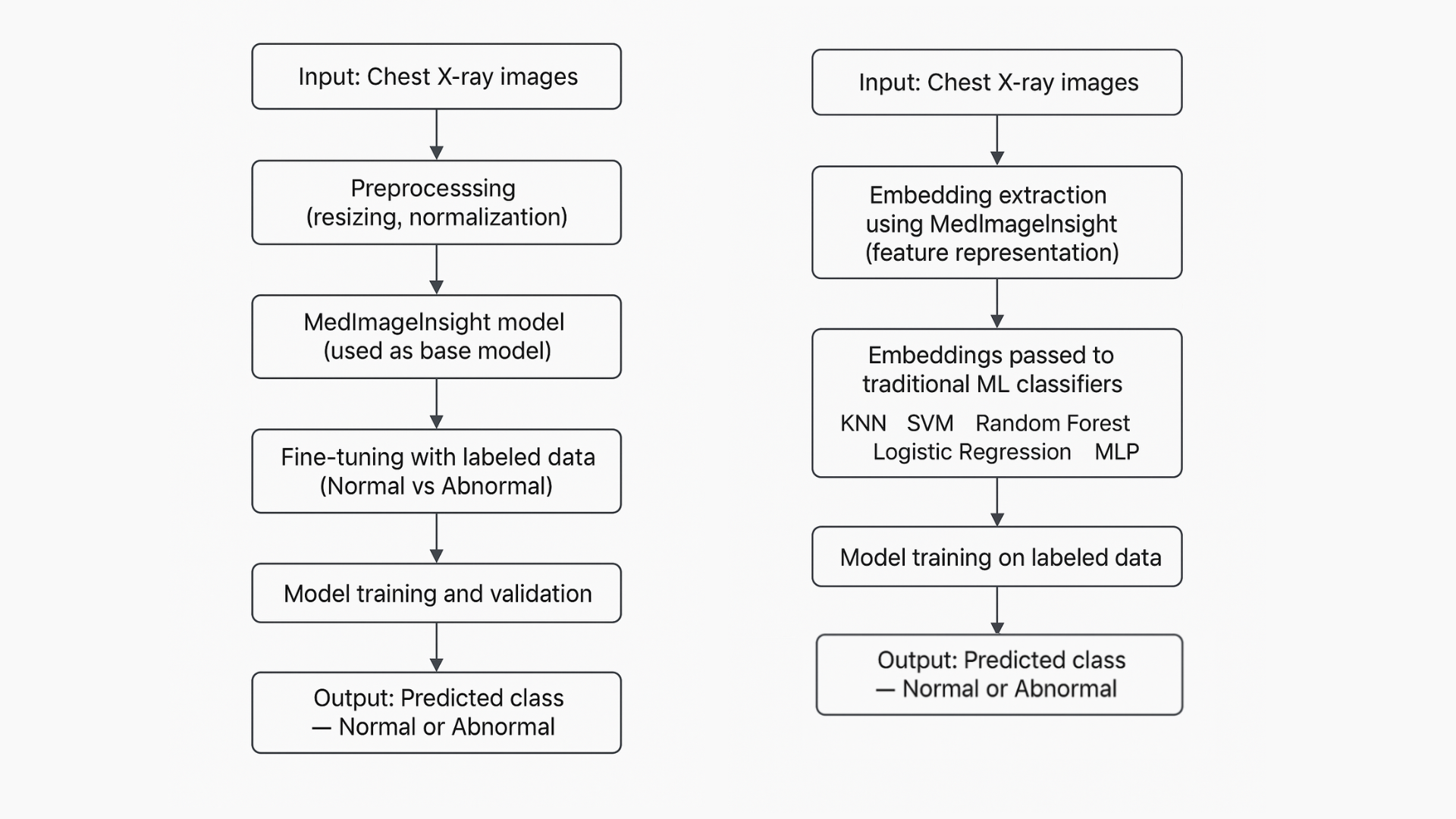}
    \caption{Workflow comparison of two model training approaches for chest X-ray image classification. (a) (on the left) End-to-end fine-tuning pipeline using the MedImageInsight model as a base network, where images undergo preprocessing before fine-tuning with labelled data (Normal vs. Abnormal). (b) (on the right) Feature extraction approach where embeddings generated by MedImageInsight are input to traditional machine learning classifiers (KNN, SVM, Random Forest, Logistic Regression, and MLP) to predict Normal vs. Abnormal.}
    \label{fig:training_workflows}
\end{figure}

Training Setup: a. For Finetuning, the training employed the AdamW optimiser with a learning rate of 1E-5 and weight decay of 0.2. The batch size was set to 48, distributed across multiple V100 GPUs, with gradient accumulation steps computed based on the batch size and number of GPUs. The learning rate followed a cosine decay schedule, with warmup steps as described in Microsoft's "Healthcare AI Examples" GitHub page [15]. In Fig. 5a, a flow diagram of the training and evaluation setup is shown.  \newline \newline
b. For transfer learning, the training process used traditional machine learning classifiers optimised via grid search with 5-fold cross-validation. Each model—K-Nearest Neighbours, Logistic Regression, Support Vector Machine, Random Forest, and Multilayer Perceptron—was trained on embeddings generated by MedImageInsight. Hyperparameter optimisation was conducted through exhaustive search over predefined parameter grids, and the best configuration was selected based on the highest cross-validated ROC-AUC score. Model training and evaluation were executed on high-performance compute instances to ensure consistency and efficiency across experiments. Fig. 5b shows a flow diagram of the training and evaluation setup for the transfer learning approach.

\subsection{Datasets}
The ChestX-ray14 dataset [13], released by the U.S. National Institutes of Health (NIH), consists of 112,120 frontal-view chest radiographs from 30,805 patients, each annotated with up to 14 thoracic disease labels derived from associated radiology reports using natural language processing. The labelled conditions include Atelectasis, Cardiomegaly, Effusion, Infiltration, Mass, Nodule, Pneumonia, Pneumothorax, Consolidation, Edema, Emphysema, Fibrosis, Pleural Thickening, and Hernia, along with a No Finding category for normal studies. All images were de-identified and released in greyscale format with variable resolutions (approximately 1024×1024 pixels). This dataset serves as a large-scale benchmark for developing and evaluating deep learning models for automated chest radiograph interpretation. \newline \newline
For Fine-tuning, we used real-world data from our partner hospitals (all patient data were de-identified before model training) in India and the UK, combined with the ChestX-ray14 dataset, to create a training dataset shown in Table 2. All 14 findings are mixed and constitute the Abnormal class, and ‘No Finding’ is renamed ‘Normal’ before training. \newline \newline
For Transfer Learning, A sample of 10000 images is taken from the data presented in Table 2. The sample is generated by categorically sampling the data, and gender is assigned to each class in equal proportions.

\begin{table}[htbp]
\centering
\scriptsize %
\caption{Distribution of chest X-ray findings across male and female patients in the dataset.}
\label{tab:gender_distribution}
\begin{tabular}{lcc}
\hline
\textbf{Category} & \textbf{Male} & \textbf{Female} \\
\hline
Atelectasis & 2468 & 2009 \\
Cardiomegaly & 766 & 860 \\
Consolidation & 1618 & 1206 \\
Edema & 730 & 671 \\
Effusion & 3025 & 2671 \\
Emphysema & 969 & 539 \\
Fibrosis & 527 & 443 \\
Hernia & 54 & 75 \\
Infiltration & 3737 & 3185 \\
Mass & 2175 & 1367 \\
No Finding & 8989 & 9011 \\
Nodule & 2178 & 1596 \\
Pleural Thickening & 1219 & 786 \\
Pneumonia & 499 & 365 \\
Pneumothorax & 1612 & 1495 \\
\hline
\end{tabular}
\end{table}

\subsection{Evaluation}
This evaluation seeks to rigorously quantify the effectiveness of MedImageInsight under two complementary paradigms: as a pretrained feature extractor within a transfer learning context, and as a fine-tuned, task-specific model optimised for chest X-ray classification.

\begin{table}[htbp]
\centering
\scriptsize %
\caption{Distribution of evaluation data across male and female patients for various chest X-ray findings.}
\label{tab:evaluation_data_distribution}
\begin{tabular}{lcc}
\hline
\textbf{Category} & \textbf{Male} & \textbf{Female} \\
\hline
Atelectasis & 823 & 645 \\
Cardiomegaly & 224 & 287 \\
Consolidation & 495 & 368 \\
Edema & 248 & 223 \\
Effusion & 1014 & 896 \\
Emphysema & 320 & 177 \\
Fibrosis & 191 & 169 \\
Hernia & 23 & 23 \\
Infiltration & 1240 & 1070 \\
Mass & 633 & 440 \\
Normal & 3058 & 2942 \\
Nodule & 758 & 523 \\
Pleural Thickening & 430 & 266 \\
Pneumonia & 158 & 103 \\
Pneumothorax & 560 & 563 \\
\hline
\end{tabular}
\end{table}

The Classifiers from both approaches were evaluated using the dataset summarised in Table 4, which contains chest X-ray images annotated for binary classification into Normal and Abnormal categories. The evaluation was conducted on a held-out test set that was not used during training to ensure unbiased performance assessment. \newline \newline
To comprehensively quantify model performance, several complementary metrics were employed. The Receiver Operating Characteristic (ROC) curve and corresponding Area Under the Curve (AUC) were used to assess the model’s discriminative ability across varying decision thresholds. Precision, Recall, and F1-Score were computed to evaluate the balance between false positives and false negatives, providing insight into classification reliability. Accuracy was used as a measure of overall correctness across all predictions. In addition, the Brier Score was calculated to assess the calibration of predicted probabilities, reflecting how closely model confidence aligned with observed outcomes. Finally, Sensitivity and Specificity were used to characterise further the model’s ability to correctly identify abnormal and normal cases, respectively—metrics particularly relevant in medical imaging contexts where diagnostic balance is critical.

\section{Discussion}
This study evaluated MedImageInsight, a medical imaging foundation model, for binary classification of chest radiographs as Normal or Abnormal using two approaches: (1) end-to-end fine-tuning and (2) transfer learning with MedImageInsight embeddings. Both methods produced clinically meaningful results, but the fine-tuned model consistently outperformed transfer learning classifiers across all metrics, including ROC-AUC, F1-score, and calibration. \newline \newline
Among the transfer learning models, the Support Vector Machine achieved the best performance, though its calibration and recall were inferior to those of the fine-tuned model. The fine-tuned classifier achieved an ROC-AUC of 0.888, comparable to leading architectures such as CheXNet [3] and Wong et al. [14], demonstrating that task-specific adaptation of foundational models effectively captures diagnostic features with limited data. \newline \newline
The improved Brier score highlights better probability calibration, a critical property for clinical decision support. Well-calibrated outputs can guide triage and prioritisation, helping radiologists manage high case volumes and reduce diagnostic delays. Furthermore, the system’s design for PACS and web-based integration supports scalable deployment in real-world workflows, particularly in resource-limited settings. \newline \newline
Nonetheless, this study is limited to binary classification and retrospective evaluation. Future work will extend the model to multi-label disease classification, prospective validation across institutions, and radiologist-in-the-loop assessments to evaluate usability and interpretability. \newline \newline
In summary, fine-tuning MedImageInsight yields robust diagnostic accuracy and superior calibration compared to transfer learning baselines, highlighting the promise of medical foundation models as scalable, high-performing tools for AI-assisted radiology.

\section{Declaration of AI-Assisted Technologies}
Portions of this manuscript were drafted and refined with the assistance of generative AI tools, including ChatGPT (OpenAI) for text generation and Grammarly for language editing. We have independently verified and edited all AI-assisted content and are accountable for the scientific accuracy and integrity of the work.

\section{Acknowledgment}
We want to express our sincere gratitude to the radiologists — Dr Sudarshan Rawat (Manipal Hospital), Dr Sapna Marda (Yashoda Hospital), Dr Srinivas Dandamudi (Aayush Hospital), Dr Sai Shravan Kumar Kosti (AIG Hospital), Dr Anjani Kumar ( Apollo Hospital), Dr Satish Babu Maddukuri (Manipal Hospital), Dr Y. Madhu Madhava Reddy (KIMS Hospital), and Dr Akarsh Yella (Vesta Diagnostics), Dr Chennakesava Gangikunta, Dr Sudheer Vinnamala (Managing Director, KIMS Hospital), Dr Pardha Sai (CMIO, KIMS Hospital), Dr Chinnababu Sunkavalli (Clinical Director-Surgical Oncology at Yashoda Hospitals and Founder of Grace Cancer Foundation) — as well as the diagnostic centres Vesta and Maha, for generously sharing their clinical expertise and supporting this research. \newline \newline
We also extend our deep appreciation to Dr P. Srinivasa Prasad (CEO, KIMS Ananthapuramu), former Director General of Railway Health Services, Ministry of Railways (India), for providing invaluable guidance and strategic insights informed by his extensive experience in national-level healthcare administration.  \newline \newline
We further acknowledge Mr S. V. Kishore Reddy (Managing Director, KIMS Saveera Hospital) for his vision, leadership, and continued encouragement toward advancing healthcare innovation and enabling this collaborative effort. \newline \newline
We also thank the Microsoft team — Mr Naiteek Sangani, Mr Javier Alvarez Valle, Dr Jameson Merkow, and Dr Noel C. F. Codella — for their technical guidance and support for the Microsoft Health Sciences platform throughout the study, as well as our colleagues at DeepInfinity Limited for their continued collaboration. \newline \newline
All collaborations were conducted in accordance with institutional ethical standards and data governance policies. No aspect of this work interfered with routine hospital operations or patient care activities.

\end{document}